\documentclass[dvips]{acta}
\usepackage{supertabular,lscape,epsfig}
\usepackage{amssymb}
\usepackage{amsmath}

\usepackage{pdflscape}
\usepackage{afterpage}

\SetPages{0}{0}

\SetVol{58}{2008}

\begin{document}

\begin{Titlepage}

\Title{An almanac of predicted microlensing events for the 21st century}

\Author{D.M. Bramich$^1$ and M.B. Nielsen$^2$}
{$^1$New York University Abu Dhabi, PO Box 129188, Saadiyat Island, Abu Dhabi, UAE\\
e-mail:dan.bramich@hotmail.co.uk\\
$^2$Center for Space Science, NYUAD Institute, New York University Abu Dhabi, PO Box 129188, Abu Dhabi, UAE\\
e-mail:mbn4@nyu.edu}

\Received{July 6th, 2018}
\end{Titlepage}

\Abstract{
Using \textit{Gaia} data release 2 (GDR2), we present an almanac of 2,509 predicted microlensing events, caused by 2,130 unique lens stars,
that will peak between 25th July 2026 and the end of the century. This work extends and completes a thorough search for future microlensing
events initiated by
Bramich (2018) and Nielsen \& Bramich (2018) using GDR2. The almanac includes 161 lenses that will cause at least two microlensing events each.
A few highlights are presented and discussed, including: (i) an astrometric microlensing event with a peak amplitude of $\sim$9.7~mas, (ii) an event that
will probe the planetary system of a lens with three known planets, and (iii) an event (resolvable from space)
where the blend of the lens and the minor source image will brighten by a detectable amount ($\sim$2~mmag) due
to the appearance of the minor source image. All of the predicted microlensing events in the almanac will exhibit astrometric
signals that are detectable by observing facilities with an angular resolution and astrometric precision similar to, or better than, that
of the \textit{Hubble Space Telescope} (e.g. NIRCam on the \textit{James Webb Space Telescope}),
although the events with the most extreme source-to-lens contrast ratios
may be challenging. Ground-based telescopes of at least 1~m in diameter can be used to observe
many of the events that are also expected to exhibit a photometric signal.
}
{gravitational lensing: micro -- methods: data analysis -- catalogs -- astrometry -- stars: fundamental parameters}

%
\section{Introduction}
\label{sec:intro}

With the recent second data release from the \textit{Gaia} satellite (GDR2; Prusti et al. 2016; Brown et al. 2018), there has been
a flurry of new work and interest on the subject of predicting
microlensing events (Kl\"uter et al. 2018; Bramich 2018; Mustill, Davies \& Lindegren 2018; Ofek 2018; Nielsen \& Bramich 2018). 
An analysis of the Tycho \textit{Gaia} Astrometric Solution (TGAS) for $\sim$2~million stars from the
first \textit{Gaia} data release (Lindegren et al. 2016) only yielded a single microlensing event prediction; namely, for the white dwarf
star LAWD~37 (McGill et al. 2018). However, GDR2 far surpasses TGAS, and any other astrometric catalogue in existence today,
in terms of its all-sky coverage, number of objects, sample volume (depth), and astrometric precision
and accuracy. With $\sim$1.7~billion objects, of which $\sim$1.3~billion have 5-parameter astrometric solutions, GDR2 can act as a catalogue
of both microlensing source and lens stars (e.g. Bramich 2018; hereafter B18), or simply as an all-sky catalogue of source stars down to $\sim$21~mag
(e.g. Nielsen \& Bramich 2018). Pre-\textit{Gaia} work on this topic has lacked sufficiently precise astrometric input
catalogues resulting in rather uncertain and unreliable predictions (e.g. Feibelman 1966; Salim \& Gould 2000;
Proft, Demleitner \& Wambsganss 2011).
Consequently, only one predicted event has been successfully detected up until now.
Sahu et al. (2017) predicted a microlensing event for Stein~2051~B (also independently predicted by
Proft, Demleitner \& Wambsganss 2011) and they acquired appropriate \textit{Hubble Space Telescope} (\textit{HST}) observations
which allowed them to measure the mass of this white dwarf as $\sim$0.675~$M_{\odot}$ with an uncertainty of $\sim$8\%.

Microlensing events can be used to directly measure the masses of single isolated stars
(e.g. Yoo et al. 2004; Yee et al. 2009; Zhu et al. 2016), for which so few measurements exist ($\lesssim$15),
and to find planetary mass companions to the lenses (e.g. Beaulieu et al. 2006).
Specifically, for bright lenses (i.e. lenses whose flux is detected),
the astrometric lensing signal alone enables the lens mass to be determined to within $\sim$1-10\% (Sahu et al. 2017).
Since directly measured stellar masses, estimated independently of the assumed internal physics, come mostly from observations of the orbital motion
of binary stars (Torres, Andersen \& Gim\'enez 2010),
astrometric microlensing provides an important channel to direct mass measurements of single stars.
Microlensing events that exhibit both photometric and astrometric signals provide even stronger constraints on the physical properties of the lens by
allowing important parameter degeneracies to be broken
(H{\o}g, Novikov \& Polnarev 1995; Miyamoto \& Yoshii 1995; Walker 1995). High-magnification photometric events
(e.g. Udalski et al. 2005) provide an excellent opportunity for exoplanet detection. The key to these microlensing applications is
to obtain good data coverage of an event from beginning-to-end to allow full characterisation. Predicting microlensing
events in advance is therefore advantageous for planning and executing the necessary observations.

The analysis of GDR2 as both a source and a lens catalogue for microlensing predictions is currently incomplete. The current
status is as follows.
Kl\"uter et al. (2018) present three cherry-picked event predictions for the stars Luyten~143-23 (two events in 2018 and 2021) and Ross~322
(one event in 2018)\footnote{We note that only the predictions for Luyten~143-23 are new. The predicted event for
Ross~322 corresponds to predicted event \#37 from Proft, Demleitner \& Wambsganss (2011).}. B18, who performed a thorough search of
GDR2 over the (extended) lifetime of the \textit{Gaia} satellite, report on 76 predicted events peaking between July 2014 and July 2026.
B18 independently find the predicted Ross~322 event from Kl\"uter et al. (2018; event ME25 in B18),
and refine the predicted LAWD~37 event from McGill et al. (2018; event ME4 in B18), thereby 
verifying the reliability of the predictions of all three groups. Furthermore, B18 present eight more
microlensing event predictions for LAWD~37 over the next decade. However, the predicted events for Luyten~143-23
from Kl\"uter et al. (2018) are not in the list of
events from B18 because this star fails the B18 lens star selection criteria on two counts. Luyten~143-23 is flagged in GDR2 as a ``duplicated source'' that
is likely to have astrometric and/or photometric problems, and also its astrometric solution suffers from highly significant excess noise
({\tt ASTROMETRIC\_EXCESS\_NOISE\_SIG}~=~18.0; dimensionless quantity in multiples of sigma).

Mustill, Davies \& Lindegren (2018) focussed on finding photometric microlensing events that will occur in the next two decades (up until July 2035).
They provide a list of 30 predicted photometric events, although only six of them are expected to reach a peak magnification
above $\sim$0.4~mmag (the bright-limit photometric precision of \textit{Gaia}). Surprisingly, none of the 30 predicted events appear in the
list of events from B18 (or in the almanac presented in this paper). A detailed inspection of the GDR2 data for the 30 lens stars reveals that all of them
exhibit highly significant excess noise in their astrometric solutions \\
(3.3~<~{\tt ASTROMETRIC\_EXCESS\_NOISE\_SIG}~<~78.4) and consequently they were rejected from the B18 lens star sample
(who adopt the requirement \\
{\tt ASTROMETRIC\_EXCESS\_NOISE\_SIG}~<~3).
Furthermore, two of the lens stars have parallax errors greater than 0.4~mas,
indicating spurious astrometric solutions (see Section~4.2.6 in B18), and another lens star is a duplicated source.
Further evidence of poor astrometric solutions for these lens stars comes from the reduced chi-squared values
of the fits, which are >2 for most of them and $\sim$17.1 in the worst case.
Considering also that the Mustill, Davies \& Lindegren (2018)
analysis should have independently found at least some of the photometric events presented in B18 and in this paper (e.g. event ME19
which will produce a photometric signal with a $\pm$1-sigma range between $\sim$0.039-0.158~mag in late 2019),
and that they did not correct for the underestimated uncertainties on the parameters of the
astrometric solutions (Brown et al. 2018), we can only conclude that their predictions should
be treated with caution and confirmed by independent calculations.

While working on the B18 paper, we realised that the astrometry in GDR2 is actually precise enough to enable
reliable microlensing predictions well past the end of any extended \textit{Gaia} mission and, in some of the more
favourable cases, up to the end of the 21st century. Hence, we have extended
the analysis from B18 to complete a thorough search of GDR2 for source-lens pairs that
will produce microlensing events in the last $\sim$75 years of this century.
The almanac of events presented in this paper corresponds specifically to bright lenses present in GDR2.
For faint lenses that do not appear in GDR2, there is still plenty of opportunity for
predicting microlensing events (e.g. Nielsen \& Bramich 2018; Ofek 2018).

The purpose and scope of this paper is to provide an almanac of predicted microlensing events
for the 21st century that can be observed with existing or planned observing facilities. For instance,
all of the predicted events in the almanac have been selected such that they will exhibit large enough
astrometric signals to be detectable by observing facilities with an angular resolution and astrometric precision
similar to, or better than, that of the \textit{HST} (White 2006; Hoffmann \& Anderson 2017),
although the events with the most extreme source-to-lens contrast ratios may be challenging.
Some of the events are also expected to exhibit a photometric signal, and in these special cases, ground-based telescopes
(typically at least 1~m in diameter) can be used for the photometric observations. The
\textit{James Webb Space Telescope}\footnote{https://jwst.nasa.gov/index.html},
due to be launched in early 2021, is suitably equipped with NIRCam (point-spread function full-width
at half-maximum of 0.064\arcs; wavelength coverage 0.6-5~$\mu$m) for the follow-up of the first events in the almanac, which
will already be rising to their peaks in the years before July 2026. The source-to-lens contrast ratio for many events
(especially those with white dwarf lenses) is greatly improved in the NIR wavebands, and coronagraphs are available
for the most difficult cases. Technology advances at such a pace that observations of the predicted events in the latter
half of the almanac will likely be routine.

The rest of this paper is organised as follows.
In Section~2, we provide a brief description of the
methods used to identify the predicted events. The details and format of the almanac are presented in
Section~3. Finally, in Section~4, we highlight some interesting predicted events
in the sample.

\section{Methods}
\label{sec:methods}

\subsection{Source-Lens Pair Selection}
\label{sec:slpselection}

\begin{table*}
\centering
\caption{List of constraints that need to be satisfied in order for a GDR2 object to be selected as a lens star.
         The symbols $\varpi$ and $\sigma[\varpi]$ represent the corrected parallax and its inflated uncertainty, respectively.
         Furthermore, $\chi^{2} = {\tt ASTROMETRIC\_CHI2\_AL}$, $\nu = {\tt ASTROMETRIC\_N\_GOOD\_OBS\_AL} - 5$,
         $L(G) = 1.2 \, \max \left\{ 1, \exp \left[ 0.2 \, (19.5 - G) \right] \right\}$, and $G = {\tt PHOT\_G\_MEAN\_MAG}$.}
\scriptsize{
\begin{tabular}{@{}lcccl@{}}
\hline
GDR2 Column Name                      & Relation & Value  & Unit & Description \\
\hline
{\tt DUPLICATED\_SOURCE}              & =        & FALSE  & -    & Reject duplicated objects that are likely to \\
                                      &          &        &      & have astrometric and photometric problems \\
{\tt FRAME\_ROTATOR\_OBJECT\_TYPE}    & =        & 0      & -    & Reject known extra-galactic objects \\
{\tt ASTROMETRIC\_PARAMS\_SOLVED}     & =        & 31     & -    & Only accept objects that have a 5-parameter \\
                                      &          &        &      & astrometric solution \\
$\varpi / \sigma[\varpi]$             & >        & 4      & -    & Only accept objects with a sufficiently \\
                                      &          &        &      & precise parallax measurement \\
$\varpi$                              & >        & 0      & mas  & Reject objects with a non-positive parallax \\
$\varpi$                              & <        & 769    & mas  & Reject objects with a parallax greater than \\
                                      &          &        &      & that of Proxima Centauri, which has a \textit{Gaia} \\
                                      &          &        &      & parallax of 768.529$\pm$0.254~mas \\
$\sigma[\varpi]$                      & <        & 0.4    & mas  & Reject objects with a spurious astrometric \\
                                      &          &        &      & solution \\
$\sqrt{\chi^{2} / \nu}$               & <        & $L(G)$ & -    & Reject objects with a spurious astrometric \\
                                      &          &        &      & solution \\
{\tt ASTROMETRIC\_EXCESS\_NOISE\_SIG} & <        & 3      & -    & Reject objects with significant excess noise \\
                                      &          &        &      & in the astrometric solution \\
{\tt PHOT\_BP\_MEAN\_MAG}             & $\neq$   & NaN    & mag  & Reject objects that do not have $G_{\mbox{\tiny BP}}$-band \\
                                      &          &        &      & photometry \\
{\tt PHOT\_RP\_MEAN\_MAG}             & $\neq$   & NaN    & mag  & Reject objects that do not have $G_{\mbox{\tiny RP}}$-band \\
                                      &          &        &      & photometry \\
\hline
\end{tabular}
}
\label{tab:lens_req}
\end{table*}

The methods used for identifying source-lens pairs from GDR2 that could potentially
lead to microlensing events follow those described in B18 with some very minor modifications.
Hence the reader is referred to B18 if further details are required. Throughout this
section and the rest of the paper, GDR2 data column names are written in
{\tt TYPEWRITER} font.

We correct the GDR2 parallaxes by adding 0.029~mas to the {\tt PARALLAX} entries
(Lindegren et al. 2018) and we inflate the uncertainties on the astrometric
parameters by 25\% (Brown et al. 2018). We then select lens stars from
GDR2 using the constraints listed in Table~1.
These constraints are the same as in B18, except that an extra requirement
that a lens star should have $G_{\mbox{\scriptsize BP}}$ and $G_{\mbox{\scriptsize RP}}$
photometry has been imposed. This avoids having to employ any external catalogues for
lens mass estimation later on. Note that the constraint $\sigma[\varpi]<0.4$~mas, which was
applied at a later stage in the data processing in B18, has been moved to the initial lens
selection in Table~1. Application of these constraints to GDR2
yields $N_{\mbox{\scriptsize L}}$=128,270,876 potential lens stars. For the source
stars, we use exactly the same sample that was selected in B18, which
consists of $N_{\mbox{\scriptsize S}}$=1,366,072,323 stars.

The angular scale-size inherent to a microlensing event is defined by the Einstein radius:
\begin{equation}
\theta_{\mbox{\scriptsize E}} = \sqrt{ \frac{4\,G M_{\mbox{\scriptsize L}}}{c^{2}} 
                                       \left( \varpi_{\mbox{\scriptsize L}} - \varpi_{\mbox{\scriptsize S}} \right) }
\label{eqn:theta_e}
\end{equation}
where $G$ is the gravitational constant, $c$ is the speed of light, $M_{\mbox{\scriptsize L}}$ is the lens mass,
and $\varpi_{\mbox{\scriptsize L}}$ and $\varpi_{\mbox{\scriptsize S}}$ are the lens and source parallaxes, respectively.
We perform the initial source-lens pair selection by computing a
conservative upper limit $\theta_{\mbox{\scriptsize E,max}}$ on the value of the Einstein radius
$\theta_{\mbox{\scriptsize E}}$ for any particular source-lens pair by assuming a maximum lens mass of
$M_{\mbox{\scriptsize L,max}}=10M_{\odot}$, a maximum lens parallax of
$\varpi_{\mbox{\scriptsize L,max}}=\varpi_{\mbox{\scriptsize L}}+3\sigma[\varpi_{\mbox{\scriptsize L}}]$
(where $\sigma[\varpi_{\mbox{\scriptsize L}}]$ is the uncertainty on the lens parallax $\varpi_{\mbox{\scriptsize L}}$),
and a source parallax of zero. This is translated into a maximum source-lens angular
separation $\theta_{\mbox{\scriptsize det}}$ within which a microlensing signal can be detected
by considering the astrometric deflections (which have a larger range of influence than
the photometric magnifications) and the best astrometric precision
achievable by any current observing facility in a single observation (excluding radio interferometry).
One obtains $\theta_{\mbox{\scriptsize det}} = \theta_{\mbox{\scriptsize E,max}}^{\,2} / 0.030$~mas,
where 0.030~mas is the bright-limit along-scan astrometric precision for \textit{Gaia}
(Rybicki et al. 2018; B18).

For each potential lens star, we calculate the value of $\theta_{\mbox{\scriptsize det}}$.
The minimum, median, and maximum values of $\theta_{\mbox{\scriptsize det}}$ over all lens
stars are $\sim$0.27, 2.69, and 826~arcsec, respectively.
To account (very) conservatively for source and lens motions, and for errors in the astrometric parameters,
we compute the following quantity for each source-lens pair:
\begin{equation}
\begin{aligned}
\theta^{\prime}_{\mbox{\scriptsize det}} = & \,\, \theta_{\mbox{\scriptsize det}}
                                             + 3\sigma[\alpha_{*,\mbox{\scriptsize ref,S}}] + 3\sigma[\alpha_{*,\mbox{\scriptsize ref,L}}]
                                             + 3\sigma[\delta_{\mbox{\scriptsize ref,S}}] + 3\sigma[\delta_{\mbox{\scriptsize ref,L}}]                                     \\
                                           & + T_{\mbox{\scriptsize rem}} \, \left( \,|\,\mu_{\alpha*,\mbox{\scriptsize S}}| + 3\sigma[\mu_{\alpha*,\mbox{\scriptsize S}}]
                                                                          + |\,\mu_{\alpha*,\mbox{\scriptsize L}}| + 3\sigma[\mu_{\alpha*,\mbox{\scriptsize L}}] \,\right) \\
                                           & + T_{\mbox{\scriptsize rem}} \, \left( \, |\,\mu_{\delta,\mbox{\scriptsize S}}| + 3\sigma[\mu_{\delta,\mbox{\scriptsize S}}]
                                                                          + |\,\mu_{\delta,\mbox{\scriptsize L}}| + 3\sigma[\mu_{\delta,\mbox{\scriptsize L}}] \,\right)   \\
                                           & + \varpi_{\mbox{\scriptsize S}} + 3\sigma[\varpi_{\mbox{\scriptsize S}}]
                                             + \varpi_{\mbox{\scriptsize L}} + 3\sigma[\varpi_{\mbox{\scriptsize L}}]                                                      \\
\label{eqn:newlim}
\end{aligned}
\end{equation}
where $\sigma[\alpha_{*,\mbox{\scriptsize ref}}] / \cos(\delta_{\mbox{\scriptsize ref}})$, $\sigma[\delta_{\mbox{\scriptsize ref}}]$,
$\sigma[\mu_{\alpha*}]$, $\sigma[\mu_{\delta}]$, and $\sigma[\varpi]$
are the uncertainties on $\alpha_{\mbox{\scriptsize ref}}$ ({\tt RA}), $\delta_{\mbox{\scriptsize ref}}$ ({\tt DEC}),
$\mu_{\alpha*}$ ({\tt PMRA}), $\mu_{\delta}$ ({\tt PMDEC}),
and $\varpi$ ({\tt PARALLAX}), respectively, and $T_{\mbox{\scriptsize rem}}=84.5$~years is
the length of time from the GDR2 reference epoch (J2015.5) until the end of the 21st century (J2100.0).
The subscripts S and L correspond to the source and the lens, respectively.
We reject all source-lens pairs for which the angular distance between them at time $t=\mbox{J}2015.5$
exceeds $\theta^{\prime}_{\mbox{\scriptsize det}}$.
This leaves 80,611,203 source-lens pairs, with 38,608,663 unique lenses, for further consideration.

We refine the source-lens pair selection as follows. For each lens star, we compute an approximate lens mass 
using the lens parallax, its mean $G$-band magnitude, the relation $L/L_{\odot}\approx(M/M_{\odot})^{4}$ for main sequence stars more massive than the Sun,
and by adopting $M_{\mbox{\scriptsize bol},\odot}\approx4.74$~mag. To conservatively account
for extinction and bolometric corrections, $M_{\mbox{\scriptsize L,max}}$ is
set to four times the lens mass estimate. For giant stars, $M_{\mbox{\scriptsize L,max}}$
is over-estimated since they are more luminous than main sequence stars, and hence the value of $M_{\mbox{\scriptsize L,max}}$ computed in this way
also serves as a maximum lens mass for giant stars.
If $M_{\mbox{\scriptsize L,max}}$ is less than the Chandrasekhar limit, then we increase it to
1.44$M_{\odot}$ to cover the possibility that the lens is a white dwarf, which also serves as an upper limit to the lens
mass for sub-solar mass main sequence stars and brown dwarfs. We then recompute $\theta_{\mbox{\scriptsize E,max}}$
for each source-lens pair using these improved upper limits on the lens masses.

We convert the new value of $\theta_{\mbox{\scriptsize E,max}}$ for each source-lens pair into an improved
maximum detection radius $\theta_{\mbox{\scriptsize det}}$ by considering the asymptotic behaviour of the 
photometric and astrometric lensing signals in both the unresolved and partially-resolved microlensing regimes
(see Section~4.2.5 in B18 for details), and by adopting the bright-limit photometric and astrometric precision of
\textit{Gaia} for a single observation (averaged over all possible scanning angles). These limits are 0.4~mmag for photometric
observations (widely applicable to ground-based telescopes and of the correct order of magnitude for space telescopes - e.g. \textit{HST})
and 0.131~mas for astrometric observations (similar to the best astrometric precision achievable by \textit{HST} of $\sim$0.2~mas,
e.g. Kains et al. 2017).
The minimum, median, and maximum values of the improved $\theta_{\mbox{\scriptsize det}}$ over all source-lens pairs
are $\sim$0.12, 0.22, and 29.6~arcsec, respectively.

For each source-lens pair, we consider their paths on the sky during the time period 25th July 2026 ($t=2461246.5$~BJD[TDB])
and 1st January 2100 ($t=2488069.5$~BJD[TDB]). The start of this time period coincides with the end of the time period considered
in B18. To compute the parallax factors for an object at any time $t$, we use tabulated values of the Solar-system barycentric coordinates
of the Earth (J2000.0 reference epoch; $X(t)$, $Y(t)$ and $Z(t)$ in au; see Section~7.2.2.3 in Urban \& Seidelmann 2013)
provided at daily intervals for the 21st century by the Jet Propulsion Laboratory HORIZONS on-line ephemeris computation
service\footnote{https://ssd.jpl.nasa.gov/horizons.cgi}. We employ cubic spline interpolation to calculate $X(t)$, $Y(t)$ and $Z(t)$
from these data for any time $t$. We reject all source-lens pairs that do not approach each other to within an angular distance
of less than $\theta_{\mbox{\scriptsize det}}$ during the adopted time period. After performing this filtering step for each source-lens pair,
we have 57,744 source-lens pairs remaining, with 53,741 unique lenses.

To filter out binary and co-moving source-lens pairs, and source-lens pairs where the lens is more distant than the source, 
we reject all source-lens pairs with $\Delta\varpi/\sigma[\Delta\varpi]<3$ where:
\begin{equation}
\frac{\Delta\varpi}{\sigma[\Delta\varpi]} = \frac{ \varpi_{\mbox{\scriptsize L}} - \varpi_{\mbox{\scriptsize S}} }
                                                 { (\sigma[\varpi_{\mbox{\scriptsize L}}]^{2} + \sigma[\varpi_{\mbox{\scriptsize S}}]^{2})^{1/2} }
\label{eqn:relparsn}
\end{equation}
We also reject all source-lens pairs for which the signal-to-noise ratio of the total relative proper motion is less than 3.
At this point we now have 43,423 source-lens pairs remaining, with 39,545 unique lenses.

Finally, with $\Delta\varpi$ guaranteed to be positive, we recompute the improved maximum detection radius
$\theta_{\mbox{\scriptsize det}}$ for the latest set of source-lens pairs using the improved upper limit
on the lens mass $M_{\mbox{\scriptsize L,max}}$ and adopting
$\Delta\varpi_{\mbox{\scriptsize max}}=\Delta\varpi+3\sigma[\Delta\varpi]$, which takes into account the
source parallax (when available). By rejecting all source-lens pairs
that do not approach each other to within an angular distance of less than $\theta_{\mbox{\scriptsize det}}$
during the adopted time period,
we are left with a final sample that consists of 42,572 source-lens pairs, with 38,717 unique lenses.

\subsection{Lens Mass Estimation}
\label{sec:lens_mass}

\begin{figure*}
\centering
\epsfig{file=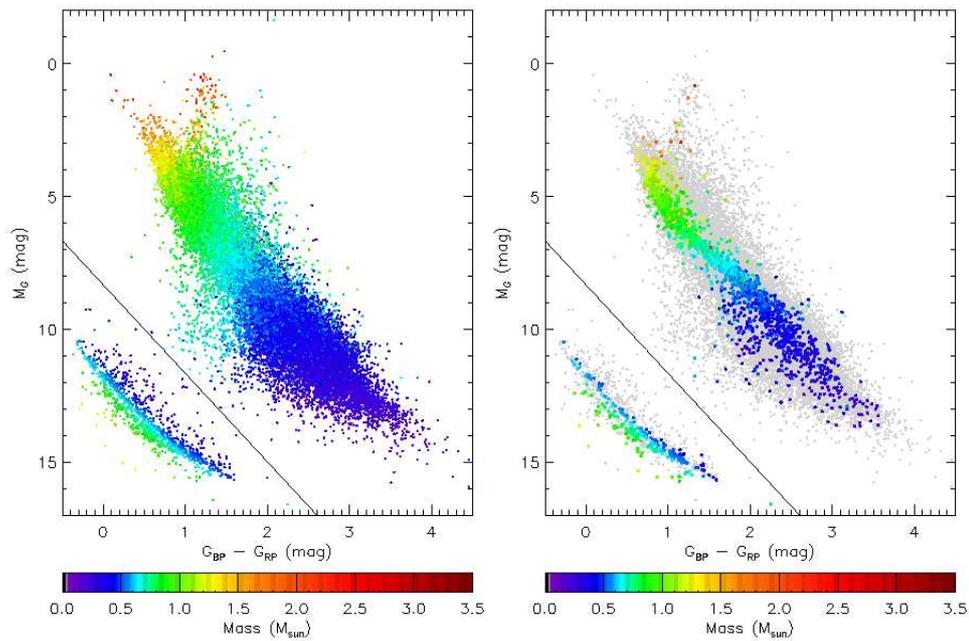,angle=0.0,width=\linewidth}
\caption{\textit{Left:} Hertzsprung-Russell diagram of $M_{G}$ versus $G_{\mbox{\scriptsize BP}}-G_{\mbox{\scriptsize RP}}$
         for the 38,717 lens stars in the final sample of source-lens pairs from Section~2.1. The continuous
         black line joining (-1,5) and (5,25) mag separates white dwarf stars from main sequence dwarfs/subdwarfs/giants
         (Kilic et al. 2018). The lens masses are indicated by the colour of the plot points (see the scale at the bottom
         of the panel).
         \textit{Right:} Same as the left-hand panel for the 2,130 lens stars in the 2,509 microlensing
         events found in Section~2.3. The points from the left-hand panel are plotted in
         the background in light grey.
         \label{fig:hrd}}
\end{figure*}

To be able to predict microlensing events and their properties from the final selection of source-lens pairs,
it is necessary to have a reasonable estimate of the lens mass in each case.
In the left-hand panel of Figure~1, we plot absolute $G$-band magnitude $M_{G}$ against
$G_{\mbox{\scriptsize BP}}-G_{\mbox{\scriptsize RP}}$ colour (using {\tt PHOT\_BP\_MEAN\_MAG} and {\tt PHOT\_RP\_MEAN\_MAG})
for the 38,717 lens stars in the final sample of source-lens pairs from Section~2.1.
The absolute $G$ magnitude is computed using:
\begin{equation}
M_{G} = G + 5\log(\varpi_{\mbox{\scriptsize L}}) + 5
\label{eqn:absgmag}
\end{equation}
where $G$ is the apparent $G$-band mean magnitude ({\tt PHOT\_G\_MEAN\_MAG}). No attempt has been made to
account for reddening and extinction in this plot. 

There are 1,007 white dwarf stars that lie below the continuous black line joining
($-$1,5) and (5,25) mag in Figure~1 (defined by Kilic et al. 2018). To estimate their masses,
we interpolate the evolutionary cooling sequences
for DA- and DB-type white dwarfs computed specifically for the \textit{Gaia} passbands (Pierre Bergeron - private communication;
Holberg \& Bergeron 2006; Kowalski \& Saumon 2006; Tremblay, Bergeron \& Gianninas 2011;
Bergeron et al. 2011). The mass estimates from the DA and DB
cooling sequences are always very similar (to within $\sim$1-15\%) and for the purposes of predicting microlensing events,
these differences are unimportant. Therefore, for each white dwarf lens star, we adopt the greater of the two mass estimates.
The lens masses estimated by this method are indicated by the colours of the plot points in Figure~1.

There are 37,710 main sequence and giant stars that lie above the continuous line in Figure~1.
We use the {\tt isochrones} Python package (Morton 2015) to estimate the lens masses, which employs the
MESA (Paxton et al. 2011, 2013, 2015) Isochrones and
Stellar Track Library (MIST; Dotter 2016; Choi et al. 2016).
For each lens star, we compute Sloan $g$ and $i$ magnitudes from $G$, $G_{\mbox{\scriptsize BP}}$ and $G_{\mbox{\scriptsize RP}}$
via the relevant transformations given in Evans et al. (2018), and we provide $\varpi_{\mbox{\scriptsize L}}$, $G$, $g$ and $i$,
along with their uncertainties, as input to {\tt isochrones}. Whenever possible, we bound-above the extinction prior in {\tt isochrones}
by using the {\tt dustmaps} Python package with the Bayestar17 three-dimensional dust maps (Green et al. 2015, 2018).
The {\tt isochrones} package maximises the posterior probability of the fundamental parameters
(mass, age, metallicity, distance, and extinction) for each lens star given the input data,
and it samples the posterior distributions using the MCMC ensemble sampler {\tt emcee} (Foreman-Mackey et al. 2013).
We use 300 walkers which each execute a burn-in of 300 steps, iterate for a further 500 steps, and the last 100 of these steps are recorded.
We adopt the median of the posterior sample as the estimate of the lens mass in each case.
Again, the lens masses estimated by this method are indicated by the colours of the plot points in Figure~1.

\subsection{Finding Microlensing Events}
\label{sec:find_events}

For each of the 42,572 source-lens pairs from Section~2.1, we perform
1,000 Monte Carlo simulations of the source and lens paths on the sky. Each simulation is generated
using the following procedure:
\begin{enumerate} 
\item We draw a set of astrometric parameters for the lens star
      from a multi-variate Gaussian distribution defined by the lens astrometric solution parameter values and their covariance
      matrix provided in GDR2. We do the same for the source star.
\item We use the lens mass estimate $M_{\mbox{\scriptsize L}}$ from Section~2.2, and the lens and source parallaxes
      $\varpi_{\mbox{\scriptsize L}}$ and $\varpi_{\mbox{\scriptsize S}}$, respectively, drawn in step~(1), to calculate
      the Einstein radius $\theta_{\mbox{\scriptsize E}}$ (Equation~1).
\item We compute the path of the source relative to the lens for the years 2020 to 2106, which fully encompasses the time period
      studied in this paper (25th July 2026 to 1st January 2100).
\item We compute the lens-to-source flux ratio $f_{\mbox{\scriptsize L}} / f_{\mbox{\scriptsize S}}$ using the \textit{Gaia} $G$-band photometry
      and we adopt \textit{Gaia}'s resolution of 103~mas (Fabricius et al. 2016). We use these values along with the relative path computed
      in step~(3) to calculate the amplitudes of the
      photometric and astrometric lensing signals in the unresolved (if necessary) and partially-resolved microlensing regimes
      (see Section~3 in this paper and Section~3 in B18).
\end{enumerate}
We then calculate the median amplitude of each of the microlensing signals over all of the simulations for
the source-lens pair. Collecting these results for the 42,572 source-lens pairs, we reject
all source-lens pairs for which none of the median microlensing signal amplitudes exceed 0.4~mmag for photometric signals or
0.131~mas for astrometric signals (Section~2.1). We also reject all source-lens pairs for which
the epoch $t_{0}$ of the microlensing event peak falls outside of the time period 25th July 2026 to 1st January 2100.
The characteristics of some of the predicted events are rather uncertain. To filter out these poorer-quality
events, we reject all source-lens pairs for which the $\pm$1-sigma range on $t_{0}$ exceeds 0.8~years.

The final set of predicted microlensing events consists of 2,509 events caused by 2,130 unique lens stars.
Eleven predicted events from B18 are present in this sample because they peak after 25th July 2026. These events
are ME8, ME9, ME14, ME33, ME41, ME43, ME49, ME61, ME62, ME67, ME73, and their predicted properties presented in this almanac
supersede those reported in B18 (e.g. ME62 has a $\sim$15.9\% probability of yielding a photometric signal of at least
$\sim$0.03 mag). We name the remaining events ME104-ME2601, following
on from the last predicted event reported in Nielsen \& Bramich (2018). The lenses consist of 213 white dwarfs and 1,917 main sequence
and giant stars. In the right-hand panel of Figure~1, we plot the Hertzsprung-Russell
diagram for the lenses in these microlensing events, with the lenses from the left-hand panel plotted in the background in
light grey.

\section{The Almanac}
\label{sec:events}

\begin{figure}
\centering
\epsfig{file=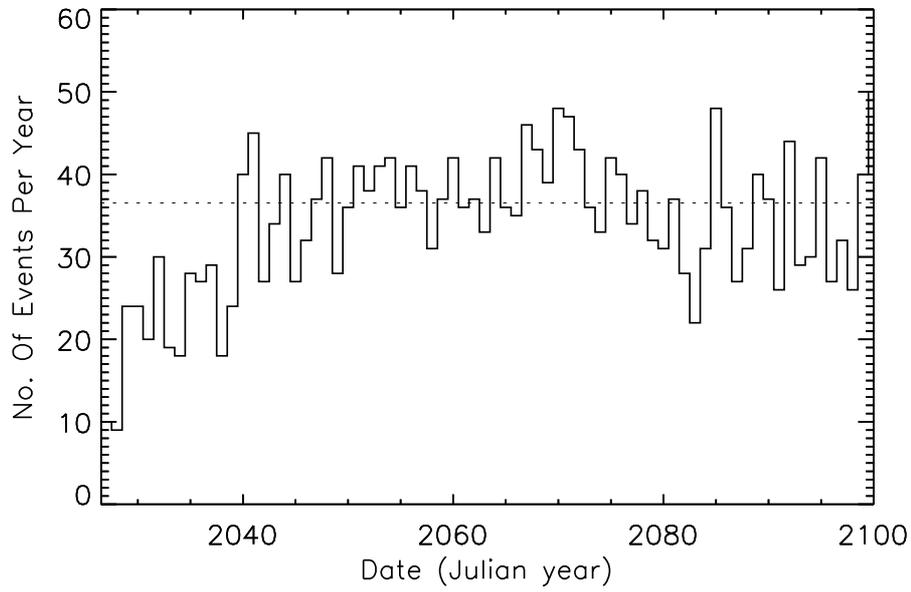,angle=0.0,width=\linewidth}
\caption{Number of predicted microlensing events per year as a function of Julian year in the almanac. The bin size is 1~year.
         The horizontal dotted line shows the mean event rate of $\sim$36.5 events per year from 2040 to 2100.
         \label{fig:t0}}
\end{figure}

\begin{figure}
\centering
\epsfig{file=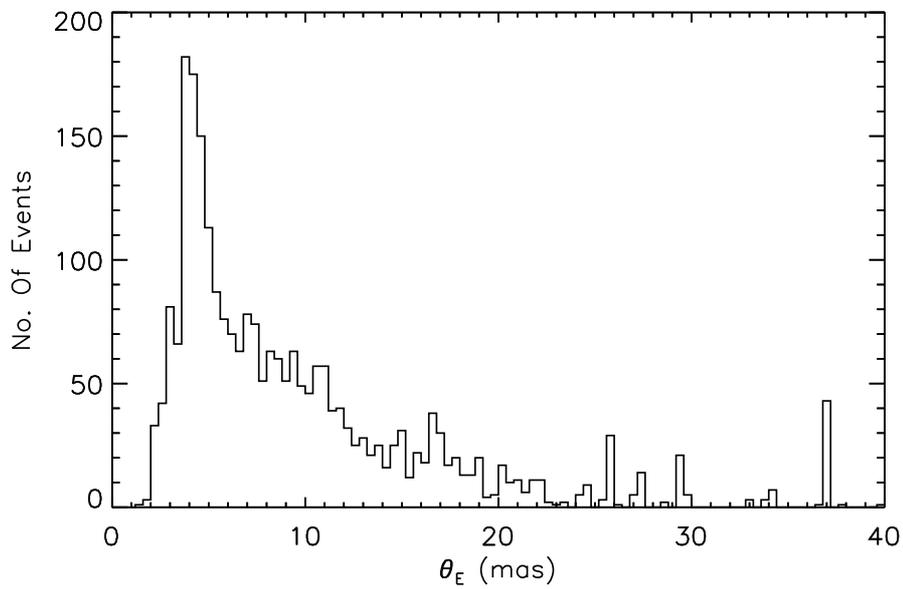,angle=0.0,width=\linewidth}
\caption{Histogram of $\theta_{\mbox{\scriptsize E}}$ for the microlensing events in the almanac. The bin size is 0.4~mas.
         \label{fig:theta_E}}
\end{figure}

\begin{figure}
\centering
\epsfig{file=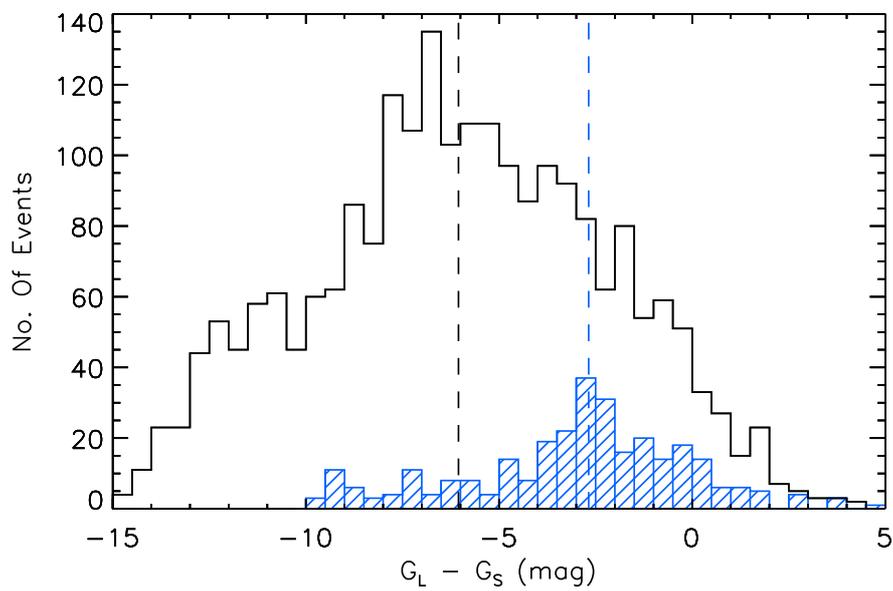,angle=0.0,width=\linewidth}
\caption{Histograms of the magnitude difference $G_{\mbox{\scriptsize L}}-G_{\mbox{\scriptsize S}}$ for the microlensing
         events in the almanac for main sequence dwarfs/subdwarfs/giants (black histogram) and white dwarfs (blue histogram
         with striped bars). The bin size is 0.5~mag. The median values for each histogram are plotted as vertical dashed lines
         ($-$6.05 and $-$2.68~mag for the black and blue histograms, respectively). 
         \label{fig:GLmGS}}
\end{figure}

The almanac of predicted microlensing events for the 21st century\footnote{Available at the Centre de Donn\'ees astronomiques de
Strasbourg: http://cdsweb.u-strasbg.fr/} is provided as a table in electronic form with 99 columns and a single row per event.
The table format and column meanings are defined in Tables~2~and~3, which also include an example row for event ME2395.

The following definitions relevant to the almanac
are reported here for convenience, and the reader is referred to Section~3 of B18 for full details. For unresolved microlensing, the observed
overall magnification $A$ of the source and lens blend is given by:
\begin{equation}
A = \frac{ u^{2} + 2 + ( f_{\mbox{\scriptsize L}} / f_{\mbox{\scriptsize S}} ) \, u \sqrt{u^{2} + 4} }
         { ( 1 + f_{\mbox{\scriptsize L}} / f_{\mbox{\scriptsize S}} ) \, u \sqrt{u^{2} + 4} } \\
\label{eqn:magnification}
\end{equation}
For angular position vectors $\vec{\phi}_{\mbox{\scriptsize S}}$ and
$\vec{\phi}_{\mbox{\scriptsize L}}$ on the celestial sphere corresponding to the source and the lens, respectively,
one may calculate the normalised source-lens separation $u$ via:
\begin{equation}
u = \left| \frac{ \vec{\phi}_{\mbox{\scriptsize S}} - \vec{\phi}_{\mbox{\scriptsize L}} }{ \theta_{\mbox{\scriptsize E}} } \right|
\label{eqn:u}
\end{equation}
The normalised source-lens separation reaches a minimum of $u=u_{0}$ at time $t=t_{0}$.
The centroid shift due to microlensing in the unresolved regime is given by:
\begin{equation}
\delta_{\mbox{\scriptsize mic}} = \frac{ \theta_{\mbox{\scriptsize E}} \, u }{ 1 + f_{\mbox{\scriptsize L}} / f_{\mbox{\scriptsize S}} }
                                         \left[ \frac{ 1 + ( f_{\mbox{\scriptsize L}} / f_{\mbox{\scriptsize S}} ) \,
                                         \left( u^{2} + 3 - u \sqrt{u^{2} + 4} \right) }
                                       { u^{2} + 2 + ( f_{\mbox{\scriptsize L}} / f_{\mbox{\scriptsize S}} ) \, u \sqrt{u^{2} + 4} } \right] \\
\label{eqn:centroidshift}
\end{equation}
In the partially-resolved microlensing regime, the major source image is resolved from the blend of the minor source image
and the lens. In this case, the major source image is magnified by $A_{1}$ relative to the source flux and it is shifted
from the nominal source position by an angular distance of $\theta_{2}$. Also, the blend of the minor source image and the lens is
magnified by $A_{\mbox{\scriptsize LI}_{2}}$ relative to the lens flux and the centroid is shifted from the nominal lens position
by an angular distance of $\theta_{\mbox{\scriptsize LI}_{2}}$. These quantities are given by:
\begin{equation}
A_{1} = \frac{ u^{2} + 2 }{ 2 u \sqrt{u^{2} + 4} } + \frac{1}{2}
\label{eqn:a1}
\end{equation}
\begin{equation}
\theta_{2} = \frac{ \theta_{\mbox{\scriptsize E}} }{2} \left( \sqrt{ u^{2} + 4 } - u \right)
\label{eqn:theta2}
\end{equation}
\begin{equation}
A_{\mbox{\scriptsize LI}_{2}} = \frac{ u^{2} + 2 + (2 ( f_{\mbox{\scriptsize L}} / f_{\mbox{\scriptsize S}} ) - 1)\, u \sqrt{u^{2} + 4} }
                                     { 2 ( f_{\mbox{\scriptsize L}} / f_{\mbox{\scriptsize S}} )\, u \sqrt{u^{2} + 4} } \\
\label{eqn:ALI2}
\end{equation}
\begin{equation}
\theta_{\mbox{\scriptsize LI}_{2}} = \theta_{\mbox{\scriptsize E}}
                                     \left[ \frac{ (u^{2} + 1) \left( \sqrt{u^{2} + 4} - u \right) - 2u }
                                                 { u^{2} + 2 + (2 ( f_{\mbox{\scriptsize L}} / f_{\mbox{\scriptsize S}} ) - 1)\, u \sqrt{u^{2} + 4} } \right] \\
\label{eqn:centroidLI2}
\end{equation}

It is important to note that in creating the almanac, no attempt has been made to identify binary stars (visual or unresolved) that may contaminate
the lens and/or source stars in some of the predicted events. The effect of binarity on the predictions for affected source-lens pairs can range from
negligible (e.g. if the orbital period is of the order of thousands of years) to invalidating a predicted event entirely (e.g. if the orbital
period is of the order of a century). However, lens and source stars in this study were selected to have astrometric solutions
that do not exhibit significant excess noise specifically to exclude the shorter period binary stars from the sample. Hence, there
should be relatively little contamination from binaries with periods of up to $\sim$10$\times$ the GDR2 time baseline (i.e. $\sim$20 years).

In Figure~2, we plot the number of predicted microlensing events per year as a function of Julian year. The event rate increases until
the year $\sim$2040, whereafter it stabilises at $\sim$36.5 events/year (dotted line). We note that source-lens pairs in GDR2
must start out as resolved objects during the period over which the GDR2 data were acquired (25th July 2014 to 23rd May 2016),
and that for GDR2, resolved source-lens pairs are separated by at least $\sim$0.5\arcs
at the \textit{Gaia} reference epoch (J2015.5; Arenou et al. 2018).
Therefore, it takes a number of years before the majority of source-lens pairs can approach each other close enough to start producing microlensing events,
resulting in the observed ramping-up of the event rate from 2015 to $\sim$2040.

In Figure~3, we plot a histogram of the Einstein radius for the microlensing events in the almanac. The largest value of
$\theta_{\mbox{\scriptsize E}}$ is $\sim$39.76~mas for the event ME675 caused by the white dwarf lens star Wolf~28 (distance $\sim$4.31~pc).
However, the majority of $\theta_{\mbox{\scriptsize E}}$ values are below $\sim$20~mas, with a peak at $\sim$4~mas. The cut-off below
$\theta_{\mbox{\scriptsize E}}\approx4$~mas is due to the fact that the more-distant slower-moving lenses do not have enough time before
the year 2100 to fully approach potential source stars. In fact, all of the lenses in the almanac are closer
than 1.9~kpc, and 95.9\% of the lenses are closer than 500~pc.

In Figure~4, we plot two histograms of the magnitude difference $G_{\mbox{\scriptsize L}}-G_{\mbox{\scriptsize S}}$ for the microlensing
events in the almanac, where $G_{\mbox{\scriptsize L}}$ and $G_{\mbox{\scriptsize S}}$ are the mean $G$-band magnitudes
for the lens and source stars, respectively. The black and blue histograms correspond to main sequence dwarfs/subdwarfs/giants and white dwarfs,
respectively. Predicted events that are most favourable for follow-up observations are those where the lens is fainter than the source.
There are 118 such events where the lens is a main sequence \\
dwarf/subdwarf/giant star and 39 such events where the lens is a white dwarf star.
The histograms, along with their median values, demonstrate that the contrast between a white dwarf lens and the source is typically
considerably less than the contrast between a main sequence dwarf/subdwarf/giant lens and the source, indicating that predicted events
with white dwarf lenses will generally be less challenging to observe.

With future \textit{Gaia} data releases (e.g. GDR3 scheduled for late 2020), we intend to refine the predictions in the almanac and to add new
predictions for stars that were not included in GDR2 (e.g. for the $\sim$17\% of stars with proper motions greater than
$\sim$0.6~arcsec~yr$^{-1}$ that are missing from GDR2 - Brown et al. 2018).

\section{A Few Highlights}
\label{sec:highlights}

\begin{figure*}
\centering
\epsfig{file=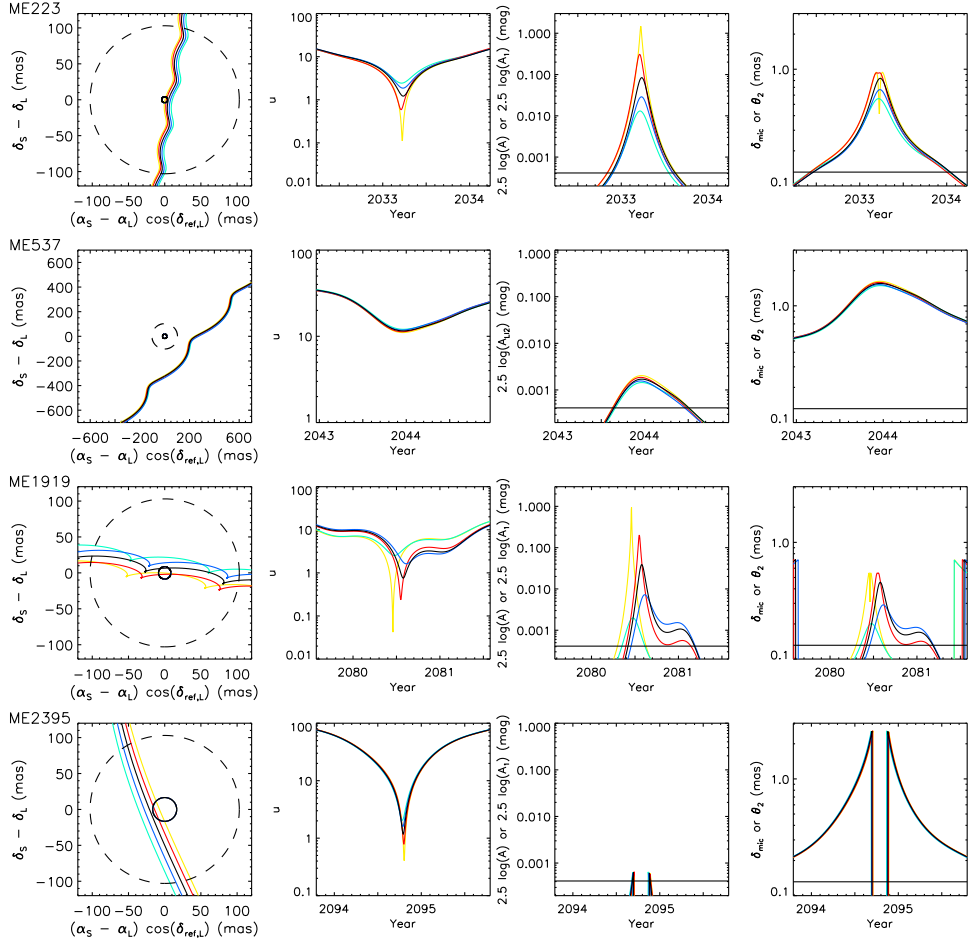,angle=0.0,width=\linewidth}
\caption{Microlensing events ME223, ME537, ME1919 and ME2395.
         In all panels, five curves are plotted with the colours yellow, red, black, blue, and cyan.
         Each curve corresponds to the 2.3, 15.9, 50, 84.1,
         and 97.7 percentiles, respectively, of the results of the Monte Carlo simulations performed
         in Section~2.3 after they have been ordered
         by increasing $u_{0}$. The yellow and cyan curves are plotted first, followed by the red and blue curves,
         and finally the black curve, which is why
         the black curve is the most visible when the individual curves are hard to distinguish.
         \textit{Left-hand panels:} Path of the source star relative to the lens star. The Einstein ring is shown as a
                                    circle of radius $\theta_{\mbox{\scriptsize E}}$
                                    centred on the lens position (also plotted five times with five different colours).
                                    The resolution of \textit{Gaia} is indicated as a circle
                                    of radius 103~mas centred on the lens position (dashed curve).
         \textit{Middle left-hand panels:} Time-evolution of the normalised source-lens separation $u$ ($\theta_{\mbox{\scriptsize E}}$).
         \textit{Middle right-hand panels:} Time-evolution of the photometric signals $2.5\log(A)$ (mag; unresolved regime) and
                                            $2.5\log(A_{1})$ (mag; partially-resolved regime).
                                            Note that for ME537, $2.5\log(A_{\mbox{\scriptsize LI}_{2}})$ (mag; partially-resolved regime)
                                            is plotted instead. Definitions of $A$, $A_{1}$ and
                                            $A_{\mbox{\scriptsize LI}_{2}}$ can be found in Section~3.
                                            The horizontal black line indicates the photometric precision limit
                                            of 0.4~mmag from Section~2.1.
         \textit{Right-hand panels:} Time-evolution of the astrometric signals $\delta_{\mbox{\scriptsize mic}}$ (mas; unresolved regime)
                                     and $\theta_{2}$ (mas; partially-resolved regime). Definitions of
                                     $\delta_{\mbox{\scriptsize mic}}$ and $\theta_{2}$ can be found in
                                     Section~3.
                                     The horizontal black line indicates the astrometric precision limit
                                     of 0.131~mas from Section~2.1.
         \label{fig:events1}}
\end{figure*}

\begin{figure*}
\centering
\epsfig{file=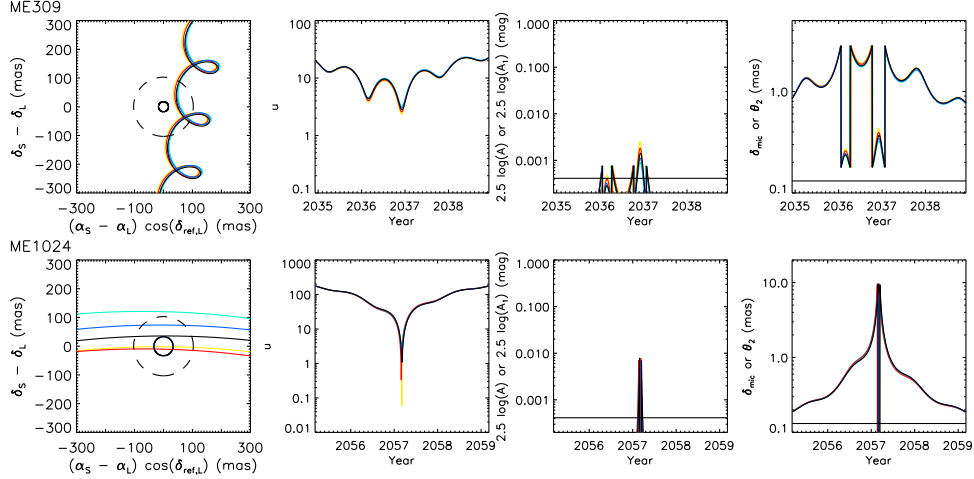,angle=0.0,width=\linewidth}
\caption{Microlensing events ME309 and ME1024. The format of the figure is the same as in Figure~5.
         \label{fig:events2}}
\end{figure*}

The almanac contains 161 lenses that will cause at least two microlensing events each, and six lenses that will cause at least 10 events each
(LAWD~37, HD~180617\footnote{This star is the primary in a visual binary (separation 74\arcs; van Biesbroeck 1961)
where the star vB~10 is the secondary component. vB~10 will also cause microlensing events in the coming years and
the relevant predictions are presented in Nielsen \& Bramich (2018).},
GJ~674, HD~39194, HD~145417 and GJ~588).
The lens that is most prolific at causing microlensing events is LAWD~37, which will produce 41 events between the years 2026 and 2100.
Interestingly, of the six lenses mentioned, GJ~674 and HD~39194 have one and three confirmed planetary companions, respectively
(Bonfils et al. 2007; Mayor et al. 2011).
Projected on the sky, these planets lie at separations of the order of $\sim$1-10~mas from their host stars.
Unfortunately, GJ~674 will not approach any source stars close enough to produce a photometric event (even though it will cause four excellent
astrometric microlensing events with source deflections greater than $\sim$3~mas). However, HD~39194
will approach a source star close enough to probe its planetary system in the event ME2395.

HD~39194 is a spectroscopically confirmed K0 dwarf star (Gray et al. 2006)
with $G_{\mbox{\scriptsize L}}\approx7.85$~mag that lies in the direction of the Large Magellanic Cloud.
The source star in event ME2395 is nearly 12 magnitudes fainter with $G_{\mbox{\scriptsize S}}\approx19.8$~mag. The event will peak on 16th October 2094
($\pm$3~d). It will achieve a maximum source deflection of $\sim$2.6~mas and it will also exhibit
a small photometric signal ($\sim$0.7~mmag). Both signals will be highly suppressed in the unresolved microlensing regime.
Figure~5 illustrates how the event will unfold.
The jumps in the photometric and astrometric curves are due to the event switching between partially-resolved and unresolved microlensing
at the \textit{Gaia} resolution of 103~mas. Clearly this behaviour is telescope/instrument dependent. The probability that the source and
the lens will approach each other to within an angular distance of less than $2\,\theta_{\mbox{\scriptsize E}}\approx33.1$~mas is 97.2\%.
This event clearly demonstrates the power of \textit{Gaia} astrometry for making reliable microlensing event predictions far into the
future ($\sim$76 years in this case), and well after the authors will have passed away.

ME223 (Figure~5) is extremely likely to exhibit an easily detectable photometric signal
($\pm$2-sigma range for $\Delta A$ is $\sim$0.013-1.657~mag)
while also yielding a strong astrometric signal ($\sim$0.84~mas). It also has the potential to become a high-magnification microlensing event
sensitive to planetary companions to the lens ($P(u_{0}<1)=0.363$). The lens star
(not in SIMBAD\footnote{http://simbad.u-strasbg.fr/simbad/}) is a main sequence
star (likely K dwarf; $\sim$0.6~$M_{\odot}$) at $\sim$253~pc with $G_{\mbox{\scriptsize L}}\approx14.62$~mag that
is only $\sim$0.62~mag brighter than the source star (at $\sim$2.4~kpc; unfortunately the source has no colour information in GDR2).
The event will peak on 28th March 2033 ($\pm$8~d).

ME537 will be a very rare event in that the blend of the lens and the minor source image will brighten by $\sim$1.7$\pm$0.2~mmag due to the
appearance of the minor source image as the event unfolds purely in the partially-resolved microlensing regime (Figure~5).
In contrast, the brightening of the source will be much smaller at $\sim$0.1~mmag. This will occur because the source star is
$\sim$3.66~mag brighter than the lens in the $G$-band ($G_{\mbox{\scriptsize L}}\approx17.44$~mag and
$G_{\mbox{\scriptsize S}}\approx13.78$~mag), the Einstein radius is reasonably large
($\theta_{\mbox{\scriptsize E}}\approx18.1$~mas), and the lens will pass close to the source
($\theta_{\mbox{\scriptsize E}} \, u_{0} \approx 209$~mas)
but the event will still remain in the partially-resolved regime (at least for some space telescopes like \textit{Gaia}
and \textit{HST}). The event will be even more favourable in redder wavebands because the lens star (white dwarf;
PM~J07228-3042) is bluer than the source star (not in SIMBAD). Apart from the photometric signal for the lens, the event will
exhibit a source deflection of $\sim$1.54~mas. The event peak will occur on 17th Dec 2043 ($\pm$3~d).
There are only two other events similar to ME537, although they are not as favourable. These events are ME1609 and ME2252.

ME1919 is an interesting event because the photometric and astrometric signals will be double-peaked (Figure~5) due to the
interplay between the lens parallax motion and its proper motion, and also because the lens will pass very
close to the source star ($P(u_{0}<1)=0.616$). The event has the potential
to achieve high-magnification, although the fact that the lens is $\sim$2.95~mag brighter than the source will
suppress the signal (redder wavebands are more favourable). The first peak will occur on 17th July 2080 ($\pm$26~d).
The lens (not in SIMBAD) is a main sequence star (likely late-K/early-M dwarf; $\sim$0.5~$M_{\odot}$).

Perhaps the funkiest event in the almanac is ME309 which will be caused
by the DA6.6 white dwarf WD~1223-659 (Gianninas, Bergeron \& Ruiz 2011).
The relative source-lens motion proceeds in loops due to the large lens parallax motion and its relatively slow proper motion.
This will result in multiple peaks in both the photometric and astrometric signals over time
(amplitudes $\sim$1~mmag and $\sim$2.8~mas, respectively; Figure~6).
The event will reach its highest two peaks during the year 2036. Unfortunately, WD~1223-659 is brighter than the source star
by $\sim$2.96~mag in the $G$-band which will suppress
the microlensing signals when the event enters the unresolved regime. In this regard, it will be better to observe the event in redder wavebands.

Finally, ME1024 is the event in the almanac that will exhibit the largest amplitude astrometric signal ($\sim$9.70$\pm$0.04~mas)
because it has
a large Einstein radius ($\theta_{\mbox{\scriptsize E}}\approx33.1$~mas) and the source is very likely to pass
within an angular distance of $2\,\theta_{\mbox{\scriptsize E}}$ from the lens
($P(u_{0}<2)=0.801$). There may also be an associated photometric signal of $\sim$8~mmag (Figure~6).
The lens star GJ~15~A is a spectroscopically confirmed M1.0 dwarf star (Trifonov et al. 2018) at a distance of $\sim$3.56~pc
that has a $\sim$5 Earth-mass planet in an $\sim$11.4~d orbit (Howard et al. 2014).
It is part of a visual binary with an M3.5 companion (GJ~15~B; L\'epine et al. 2013) that has an orbital period of
$\sim$1250 years (Romanenko \& Kiselev 2014). Since the peak of the event will occur in $\sim$39~years time on 7th March 2057 ($\pm$7~d),
the binary motion may somewhat affect the details of the event prediction presented here. While the photometric event
may or may not occur as a consequence, the amplitude of the astrometric signal is unlikely to change much since
it is much less dependent on a precise source-lens alignment. In the unresolved microlensing regime, the lens star
will highly suppress the microlensing signals since it is $\sim$13.5~mag brighter than the source
($G_{\mbox{\scriptsize L}}\approx7.22$~mag and $G_{\mbox{\scriptsize S}}\approx20.75$~mag).
The almanac contains various similar very high amplitude astrometric microlensing events.

The predicted events highlighted in this section illustrate how the current and future versions of the almanac have
the potential to provide an invaluable resource for microlensing studies.


\Acknow{
We thank the anonymous referee for taking the time to review this paper and for providing
insightful comments that served to improve the content.
DMB acknowledges the support of the NYU Abu Dhabi Research Enhancement Fund under grant RE124.
MBN is supported by NYUAD Institute grant G1502.
Many of the calculations performed in this paper employed code from the {\tt DanIDL} library of {\tt IDL}
routines (Bramich 2017) available at \\
http://www.danidl.co.uk. We are
grateful to Pierre Bergeron for providing the white dwarf cooling models in the \textit{Gaia} passbands.
This research was carried out on the High Performance Computing resources at New York University Abu Dhabi.
Thanks goes to Nasser Al Ansari, Muataz Al Barwani, Guowei He and Fayizal Mohammed Kunhi for their excellent
support. We would also like to thank the Solar System Dynamics Group at the Jet Propulsion Laboratory
for providing the Horizons On-Line Ephemeris System.
We made extensive use of the SIMBAD and VizieR web-resources as provided by
the Centre de Donn\'ees astronomiques de Strasbourg.
This work has made use of data from the European Space Agency (ESA) mission
\textit{Gaia} (https://www.cosmos.esa.int/gaia), processed by the \textit{Gaia}
Data Processing and Analysis Consortium (DPAC, \\
https://www.cosmos.esa.int/web/gaia/dpac/consortium). Funding for the DPAC
has been provided by national institutions, in particular the institutions
participating in the \textit{Gaia} Multilateral Agreement.
}



\afterpage{\clearpage}

\begin{table*}
\centering
\caption{Almanac of predicted microlensing events with peaks occurring between 25th July 2026 ($t=2461246.5$~BJD[TDB])
         and 1st January 2100 ($t=2488069.5$~BJD[TDB]). Events are ordered by increasing median epoch of the peak $t_{0}$ (column 63). The values in columns
         38-99 are computed from the 1,000 Monte Carlo simulations performed for each source-lens pair in Section~2.3.
        }
\scriptsize{
\begin{tabular}{@{}l@{\hspace{4.5pt}}l@{\hspace{4.5pt}}c@{\hspace{4.5pt}}l@{}}
\hline
Column & Column                                  & Example             & Description \\
Number & Name                                    & Event ME2395        &             \\
\hline
1      & Event Name                              & ME2395              & The name of the predicted microlensing event                \\
2      & Lens Spectral Type                      & MS                  & Lens star type: WD = white dwarf, MS = main                 \\
       &                                         &                     & sequence/subdwarf/giant                                     \\
3      & Lens GDR2 ID                            & 4657193606465368704 & Lens GDR2 source ID                                         \\
4      & $\alpha_{\mbox{\tiny ref,L}}$           & 86.1290710186       & Lens right ascension at the reference epoch J2015.5 (deg)   \\
5      & $\sigma[\alpha_{*,\mbox{\tiny ref,L}}]$ & 0.042               & Uncertainty on $\alpha_{\mbox{\tiny ref,L}} \cos(\delta_{\mbox{\tiny ref,L}})$ (mas) \\
6      & $\delta_{\mbox{\tiny ref,L}}$           & $-$70.1382377141    & Lens declination at the reference epoch J2015.5 (deg)       \\
7      & $\sigma[\delta_{\mbox{\tiny ref,L}}]$   & 0.048               & Uncertainty on $\delta_{\mbox{\tiny ref,L}}$ (mas)          \\
8      & $\mu_{\alpha*,\mbox{\tiny L}}$          & $-$309.424          & Lens proper motion in right ascension (mas/year)            \\
9      & $\sigma[\mu_{\alpha*,\mbox{\tiny L}}]$  & 0.078               & Uncertainty on $\mu_{\alpha*,\mbox{\tiny L}}$ (mas/year)    \\
10     & $\mu_{\delta,\mbox{\tiny L}}$           & 1238.780            & Lens proper motion in declination (mas/year)                \\
11     & $\sigma[\mu_{\delta,\mbox{\tiny L}}]$   & 0.098               & Uncertainty on $\mu_{\delta,\mbox{\tiny L}}$ (mas/year)     \\
12     & $\varpi_{\mbox{\tiny L}}$               & 37.860              & Lens parallax (mas)                                         \\
13     & $\sigma[\varpi_{\mbox{\tiny L}}]$       & 0.047               & Uncertainty on $\varpi_{\mbox{\tiny L}}$ (mas)              \\
14     & $G_{\mbox{\tiny L}}$                    & 7.8520              & Lens $G$-band mean magnitude (mag)                          \\
15     & $\sigma[G_{\mbox{\tiny L}}]$            & 0.0002              & Uncertainty on $G_{\mbox{\tiny L}}$ (mag)                   \\
16     & $G_{\mbox{\tiny BP,L}}$                 & 8.2816              & Lens $G_{\mbox{\tiny BP}}$-band mean magnitude (mag)        \\
17     & $\sigma[G_{\mbox{\tiny BP,L}}]$         & 0.0016              & Uncertainty on $G_{\mbox{\tiny BP,L}}$ (mag)                \\
18     & $G_{\mbox{\tiny RP,L}}$                 & 7.3085              & Lens $G_{\mbox{\tiny RP}}$-band mean magnitude (mag)        \\
19     & $\sigma[G_{\mbox{\tiny RP,L}}]$         & 0.0039              & Uncertainty on $G_{\mbox{\tiny RP,L}}$ (mag)                \\
20     & $M_{\mbox{\tiny L}}$                    & 0.89                & Lens mass estimate ($M_{\odot}$)                            \\
21     & Source GDR2 ID                          & 4657195114116751360 & Source GDR2 source ID                                       \\
22     & $\alpha_{\mbox{\tiny ref,S}}$           & 86.1090246836       & Source right ascension at the reference epoch J2015.5 (deg) \\
23     & $\sigma[\alpha_{*,\mbox{\tiny ref,S}}]$ & 0.904               & Uncertainty on $\alpha_{\mbox{\tiny ref,S}} \cos(\delta_{\mbox{\tiny ref,S}})$ (mas) \\
24     & $\delta_{\mbox{\tiny ref,S}}$           & $-$70.1109601904    & Source declination at the reference epoch J2015.5 (deg)     \\
25     & $\sigma[\delta_{\mbox{\tiny ref,S}}]$   & 1.194               & Uncertainty on $\delta_{\mbox{\tiny ref,S}}$ (mas)          \\
26     & $\mu_{\alpha*,\mbox{\tiny S}}$          & -                   & Source proper motion in right ascension (mas/year)          \\
27     & $\sigma[\mu_{\alpha*,\mbox{\tiny S}}]$  & -                   & Uncertainty on $\mu_{\alpha*,\mbox{\tiny S}}$ (mas/year)    \\
28     & $\mu_{\delta,\mbox{\tiny S}}$           & -                   & Source proper motion in declination (mas/year)              \\
29     & $\sigma[\mu_{\delta,\mbox{\tiny S}}]$   & -                   & Uncertainty on $\mu_{\delta,\mbox{\tiny S}}$ (mas/year)     \\
30     & $\varpi_{\mbox{\tiny S}}$               & -                   & Source parallax (mas)                                       \\
31     & $\sigma[\varpi_{\mbox{\tiny S}}]$       & -                   & Uncertainty on $\varpi_{\mbox{\tiny S}}$ (mas)              \\
32     & $G_{\mbox{\tiny S}}$                    & 19.7920             & Source $G$-band mean magnitude (mag)                        \\
33     & $\sigma[G_{\mbox{\tiny S}}]$            & 0.0094              & Uncertainty on $G_{\mbox{\tiny S}}$ (mag)                   \\
34     & $G_{\mbox{\tiny BP,S}}$                 & -                   & Source $G_{\mbox{\tiny BP}}$-band mean magnitude (mag)      \\
35     & $\sigma[G_{\mbox{\tiny BP,S}}]$         & -                   & Uncertainty on $G_{\mbox{\tiny BP,S}}$ (mag)                \\
36     & $G_{\mbox{\tiny RP,S}}$                 & -                   & Source $G_{\mbox{\tiny RP}}$-band mean magnitude (mag)      \\
37     & $\sigma[G_{\mbox{\tiny RP,S}}]$         & -                   & Uncertainty on $G_{\mbox{\tiny RP,S}}$ (mag)                \\
38     & 2.3\%tile $\theta_{\mbox{\tiny E}}$     & 16.527              & 2.3 percentile of the Einstein radius (mas)                 \\
39     & 15.9\%tile $\theta_{\mbox{\tiny E}}$    & 16.537              & 15.9 percentile of the Einstein radius (mas)                \\
40     & Median $\theta_{\mbox{\tiny E}}$        & 16.547              & Median of the Einstein radius (mas)                         \\
41     & 84.1\%tile $\theta_{\mbox{\tiny E}}$    & 16.557              & 84.1 percentile of the Einstein radius (mas)                \\
42     & 97.7\%tile $\theta_{\mbox{\tiny E}}$    & 16.567              & 97.7 percentile of the Einstein radius (mas)                \\
43     & 2.3\%tile $u_{0}$                       & 0.40                & 2.3 percentile of the minimum normalised source-lens        \\
       &                                         &                     & separation $u_{0}$ ($\theta_{\mbox{\tiny E}}$)              \\
44     & 15.9\%tile $u_{0}$                      & 0.79                & 15.9 percentile of the minimum normalised source-lens       \\
       &                                         &                     & separation $u_{0}$ ($\theta_{\mbox{\tiny E}}$)              \\
45     & Median $u_{0}$                          & 1.18                & Median of the minimum normalised source-lens                \\
       &                                         &                     & separation $u_{0}$ ($\theta_{\mbox{\tiny E}}$)              \\
46     & 84.1\%tile $u_{0}$                      & 1.59                & 84.1 percentile of the minimum normalised source-lens       \\
       &                                         &                     & separation $u_{0}$ ($\theta_{\mbox{\tiny E}}$)              \\
47     & 97.7\%tile $u_{0}$                      & 2.05                & 97.7 percentile of the minimum normalised source-lens       \\
       &                                         &                     & separation $u_{0}$ ($\theta_{\mbox{\tiny E}}$)              \\
48     & $P(u_{0}<1)$                            & 0.338               & Probability that the event will have $u_{0}<1$              \\
49     & $P(u_{0}<2)$                            & 0.972               & Probability that the event will have $u_{0}<2$              \\
50     & $P(u_{0}<5)$                            & 1.000               & Probability that the event will have $u_{0}<5$              \\
51     & $P(u_{0}<10)$                           & 1.000               & Probability that the event will have $u_{0}<10$             \\
\hline
\end{tabular}
}
\label{tab:almanac}
\end{table*}

\afterpage{\clearpage}

\begin{table*}
\centering
\caption{Table 2 continued. Definitions of $A$, $A_{1}$, $A_{\mbox{\scriptsize LI}_{2}}$, $\delta_{\mbox{\scriptsize mic}}$, $\theta_{2}$ and
         $\theta_{\mbox{\scriptsize LI}_{2}}$ can be found in Section~3.}
\scriptsize{
\begin{tabular}{@{}l@{\hspace{4.5pt}}l@{\hspace{4.5pt}}c@{\hspace{4.5pt}}l@{}}
\hline
Column & Column                                     & Example                 & Description \\
Number & Name                                       & Event ME2395            &             \\
\hline
52     & 2.3\%tile $\theta_{\mbox{\tiny E}} \, u_{0}$ & 6.67                  & 2.3 percentile of the minimum source-lens separation                      \\
       &                                            &                         & $\theta_{\mbox{\tiny E}} \, u_{0}$ (mas)                                  \\
53     & 15.9\%tile $\theta_{\mbox{\tiny E}} \, u_{0}$ & 13.00                & 15.9 percentile of the minimum source-lens separation                     \\
       &                                            &                         & $\theta_{\mbox{\tiny E}} \, u_{0}$ (mas)                                  \\
54     & Median $\theta_{\mbox{\tiny E}} \, u_{0}$  & 19.47                   & Median of the minimum source-lens separation $\theta_{\mbox{\tiny E}} \, u_{0}$ (mas) \\
55     & 84.1\%tile $\theta_{\mbox{\tiny E}} \, u_{0}$ & 26.30                & 84.1 percentile of the minimum source-lens separation                     \\
       &                                            &                         & $\theta_{\mbox{\tiny E}} \, u_{0}$ (mas)                                  \\
56     & 97.7\%tile $\theta_{\mbox{\tiny E}} \, u_{0}$ & 33.83                & 97.7 percentile of the minimum source-lens separation                     \\
       &                                            &                         & $\theta_{\mbox{\tiny E}} \, u_{0}$ (mas)                                  \\
57     & $P(\theta_{\mbox{\tiny E}} \, u_{0}<100~\mbox{mas})$  & 1.000        & Probability that the event will have $\theta_{\mbox{\tiny E}} \, u_{0}<100$~mas  \\
58     & $P(\theta_{\mbox{\tiny E}} \, u_{0}<200~\mbox{mas})$  & 1.000        & Probability that the event will have $\theta_{\mbox{\tiny E}} \, u_{0}<200$~mas  \\
59     & $P(\theta_{\mbox{\tiny E}} \, u_{0}<500~\mbox{mas})$  & 1.000        & Probability that the event will have $\theta_{\mbox{\tiny E}} \, u_{0}<500$~mas  \\
60     & $P(\theta_{\mbox{\tiny E}} \, u_{0}<1000~\mbox{mas})$ & 1.000        & Probability that the event will have $\theta_{\mbox{\tiny E}} \, u_{0}<1000$~mas \\
61     & 2.3\%tile $t_{0}$                          & 2094.77563              & 2.3 percentile of the epoch of the event peak $t_{0}$ (Julian year)       \\
62     & 15.9\%tile $t_{0}$                         & 2094.78296              & 15.9 percentile of the epoch of the event peak $t_{0}$ (Julian year)      \\
63     & Median $t_{0}$                             & 2094.78912              & Median of the epoch of the event peak $t_{0}$ (Julian year)               \\
64     & 84.1\%tile $t_{0}$                         & 2094.79582              & 84.1 percentile of the epoch of the event peak $t_{0}$ (Julian year)      \\
65     & 97.7\%tile $t_{0}$                         & 2094.80240              & 97.7 percentile of the epoch of the event peak $t_{0}$ (Julian year)      \\
66     & 2.3\%tile $\Delta(A,A_{1})$                & 0.0007                  & 2.3 percentile of the difference between the minimum and                  \\
       &                                            &                         & maximum source magnifications over the time period                        \\
       &                                            &                         & adopted in this paper (mag)                                               \\
67     & 15.9\%tile $\Delta(A,A_{1})$               & 0.0007                  & 15.9 percentile of $\Delta(A,A_{1})$ (mag)                                \\
68     & Median $\Delta(A,A_{1})$                   & 0.0007                  & Median of $\Delta(A,A_{1})$ (mag)                                         \\
69     & 84.1\%tile $\Delta(A,A_{1})$               & 0.0007                  & 84.1 percentile of $\Delta(A,A_{1})$ (mag)                                \\
70     & 97.7\%tile $\Delta(A,A_{1})$               & 0.0007                  & 97.7 percentile of $\Delta(A,A_{1})$ (mag)                                \\
71     & 2.3\%tile $T[\Delta(A,A_{1})]$             & 13.07                   & 2.3 percentile of the amount of time that the source spends               \\
       &                                            &                         & with its magnification above $\min \{ A,A_{1} \} + \Delta(A,A_{1})/2$ (d) \\
72     & 15.9\%tile $T[\Delta(A,A_{1})]$            & 13.20                   & 15.9 percentile of $T[\Delta(A,A_{1})]$ (d)                               \\
73     & Median $T[\Delta(A,A_{1})]$                & 13.37                   & Median of $T[\Delta(A,A_{1})]$ (d)                                        \\
74     & 84.1\%tile $T[\Delta(A,A_{1})]$            & 13.62                   & 84.1 percentile of $T[\Delta(A,A_{1})]$ (d)                               \\
75     & 97.7\%tile $T[\Delta(A,A_{1})]$            & 13.95                   & 97.7 percentile of $T[\Delta(A,A_{1})]$ (d)                               \\
76     & $P(\Delta(A,A_{1})>0.4~\mbox{mmag})$       & 1.000                   & Probability that the event will yield a source magnification              \\
       &                                            &                         & above 0.4~mmag                                                            \\
77     & 2.3\%tile $\Delta A_{\mbox{\tiny LI}_{2}}$ & 0.0000                  & 2.3 percentile of the difference between the minimum and                  \\
       &                                            &                         & maximum lens magnifications over the time period                          \\
       &                                            &                         & adopted in this paper (mag)                                               \\
78     & 15.9\%tile $\Delta A_{\mbox{\tiny LI}_{2}}$ & 0.0000                 & 15.9 percentile of $\Delta A_{\mbox{\tiny LI}_{2}}$ (mag)                 \\
79     & Median $\Delta A_{\mbox{\tiny LI}_{2}}$    & 0.0000                  & Median of $\Delta A_{\mbox{\tiny LI}_{2}}$ (mag)                          \\
80     & 84.1\%tile $\Delta A_{\mbox{\tiny LI}_{2}}$ & 0.0000                 & 84.1 percentile of $\Delta A_{\mbox{\tiny LI}_{2}}$ (mag)                 \\
81     & 97.7\%tile $\Delta A_{\mbox{\tiny LI}_{2}}$ & 0.0000                 & 97.7 percentile of $\Delta A_{\mbox{\tiny LI}_{2}}$ (mag)                 \\
82     & $P(\Delta A_{\mbox{\tiny LI}_{2}}>0.4~\mbox{mmag})$ & 0.000          & Probability that the event will yield a lens magnification                \\
       &                                            &                         & above 0.4~mmag                                                            \\
83     & 2.3\%tile $\Delta(\delta_{\mbox{\tiny mic}},\theta_{2})$ & 2.586     & 2.3 percentile of the difference between the minimum and                  \\
       &                                            &                         & maximum source astrometric shifts over the time period                    \\
       &                                            &                         & adopted in this paper (mas)                                               \\
84     & 15.9\%tile $\Delta(\delta_{\mbox{\tiny mic}},\theta_{2})$ & 2.589    & 15.9 percentile of $\Delta(\delta_{\mbox{\tiny mic}},\theta_{2})$ (mas)   \\
85     & Median $\Delta(\delta_{\mbox{\tiny mic}},\theta_{2})$ & 2.592        & Median of $\Delta(\delta_{\mbox{\tiny mic}},\theta_{2})$ (mas)            \\
86     & 84.1\%tile $\Delta(\delta_{\mbox{\tiny mic}},\theta_{2})$ & 2.595    & 84.1 percentile of $\Delta(\delta_{\mbox{\tiny mic}},\theta_{2})$ (mas)   \\
87     & 97.7\%tile $\Delta(\delta_{\mbox{\tiny mic}},\theta_{2})$ & 2.598    & 97.7 percentile of $\Delta(\delta_{\mbox{\tiny mic}},\theta_{2})$ (mas)   \\
88     & 2.3\%tile $T[\Delta(\delta_{\mbox{\tiny mic}},\theta_{2})]$ & 67.22  & 2.3 percentile of the amount of time that the source spends               \\
       &                                                             &        & with its astrometric shift above                                          \\
       &                                                             &        & $\min \{ \delta_{\mbox{\tiny mic}},\theta_{2} \} + \Delta(\delta_{\mbox{\tiny mic}},\theta_{2})/2$ (d) \\
89     & 15.9\%tile $T[\Delta(\delta_{\mbox{\tiny mic}},\theta_{2})]$ & 67.62 & 15.9 percentile of $T[\Delta(\delta_{\mbox{\tiny mic}},\theta_{2})]$ (d)  \\
90     & Median $T[\Delta(\delta_{\mbox{\tiny mic}},\theta_{2})]$     & 68.20 & Median of $T[\Delta(\delta_{\mbox{\tiny mic}},\theta_{2})]$ (d)           \\
91     & 84.1\%tile $T[\Delta(\delta_{\mbox{\tiny mic}},\theta_{2})]$ & 68.97 & 84.1 percentile of $T[\Delta(\delta_{\mbox{\tiny mic}},\theta_{2})]$ (d)  \\
92     & 97.7\%tile $T[\Delta(\delta_{\mbox{\tiny mic}},\theta_{2})]$ & 70.05 & 97.7 percentile of $T[\Delta(\delta_{\mbox{\tiny mic}},\theta_{2})]$ (d)  \\
93     & $P(\Delta(\delta_{\mbox{\tiny mic}},\theta_{2})>0.131~\mbox{mas})$ & 1.000 & Probability that the event will yield a source astrometric          \\
       &                                                  &                   & shift above 0.131~mas                                                     \\
94     & 2.3\%tile $\Delta \theta_{\mbox{\tiny LI}_{2}}$  & 0.000             & 2.3 percentile of the difference between the minimum and                  \\
       &                                            &                         & maximum lens astrometric shifts over the time period                      \\
       &                                            &                         & adopted in this paper (mas)                                               \\
95     & 15.9\%tile $\Delta \theta_{\mbox{\tiny LI}_{2}}$ & 0.000             & 15.9 percentile of $\Delta \theta_{\mbox{\tiny LI}_{2}}$ (mas)            \\
96     & Median $\Delta \theta_{\mbox{\tiny LI}_{2}}$     & 0.000             & Median of $\Delta \theta_{\mbox{\tiny LI}_{2}}$ (mas)                     \\
97     & 84.1\%tile $\Delta \theta_{\mbox{\tiny LI}_{2}}$ & 0.000             & 84.1 percentile of $\Delta \theta_{\mbox{\tiny LI}_{2}}$ (mas)            \\
98     & 97.7\%tile $\Delta \theta_{\mbox{\tiny LI}_{2}}$ & 0.000             & 97.7 percentile of $\Delta \theta_{\mbox{\tiny LI}_{2}}$ (mas)            \\
99     & $P(\Delta \theta_{\mbox{\tiny LI}_{2}}>0.131~\mbox{mas})$ & 0.000    & Probability that the event will yield a lens astrometric                  \\
       &                                            &                         & shift above 0.131~mas                                                     \\
\hline
\end{tabular}
}
\label{tab:almanac2}
\end{table*}

\end{document}